\documentclass[11pt]{article}
\usepackage{graphicx}

\begin{document}
\setlength{\baselineskip}{0.18in}
\newcommand{\be}{\begin{eqnarray}}
\newcommand{\ee}{\end{eqnarray}}
\newcommand{\bi}{\bibitem}
\newcommand{\lar}{\leftarrow}
\newcommand{\rar}{\rightarrow}
\newcommand{\lrar}{\leftrightarrow}
\newcommand{\mpl}{m_{Pl}^2}
\newcommand{\mplq}{m_{Pl}}
\newcommand{\rmn}{R_{\mu\nu}}
\newcommand{\gmn}{g_{\mu\nu}}
\newcommand{\vacr}{\rho_{vac}}

\begin{center}
\vglue .06in
{\Large \bf { 
Stability of a cosmological model with dynamical cancellation of vacuum 
energy
  }
}
\bigskip
\\{\bf A.D. Dolgov}$^{(a)(b)(c)}$ and 
{\bf M. Kawasaki}$^{(c)}$
\\[.2in]
$^{(a)}${\it INFN, sezione di Ferrara,
Via Paradiso, 12 - 44100 Ferrara,
Italy} \\
$^{(b)}${\it ITEP, Bol. Cheremushkinskaya 25, Moscow 113259, Russia.
}  \\
$^{(c)}${\it Research Center for the Early Universe, Graduate School of 
Science, \\University of Tokyo, Tokyo 113-0033, Japan
}

\end{center}

\vspace{.3in}
\begin{abstract}
The stability of cosmological solutions in the recently suggested 
specific mechanism of dynamical compensation of vacuum energy is 
studied. It is found that the solutions in the original version 
lead to cosmological singularity which could be reached in final 
(and short) time. A modification of the interaction of the 
compensating field with gravity is suggested which allows to escape
such singularity. It is shown that generic cosmological solution in 
this model tends to the Friedmann expansion regime even starting
from initially large vacuum energy.
\end{abstract}

\bigskip

In a recent paper~\cite{ad-mk-03} we considered a mechanism 
of gross reduction of vacuum energy, $\rho_{vac}$, by a scalar field, 
$\phi$, coupled to gravity in non-minimal and rather unusual way 
(see below Eq.~(\ref{A})). The energy of the condensate of this 
field in de Sitter background could 
diminish $\vacr$ down to the cosmological critical
energy density, $\rho_c (t) \sim \mpl/t^2$, transforming exponential
cosmological expansion into the usual Friedmann one. The resulting 
equation of state becomes different from the vacuum one, $p=-\rho$.
The idea of dynamical adjustment of vacuum energy is quite 
old~\cite{dolgov82} but no convincing and realistic mechanism had 
been found for a long time. Though lacking a realistic
model, the mechanism of adjustment has two impressive and
attractive features: first,
the compensation of vacuum energy with the desired accuracy of about 
100 orders of magnitude and, second, not complete compensation but only
down to time dependent remnant, $\sim \mpl/t^2$. The latter was in fact
a generic prediction of adjustment models~\cite{dolgov82,dolgov85} 
long before cosmological dark energy had been discovered~\cite{dark-en}.
For a review of the problem of vacuum energy, see e.g. 
Refs.~\cite{rev-lam}. 

However, the concrete models of adjustment discussed in the 
literature~\cite{dolgov82,dolgov85,adjust-all,mr03}
suffer from numerous shortcomings and realistic cosmology
which includes dynamical adjustment of vacuum energy has yet to 
be found. It seems clear that vacuum and dark energies are surely 
related and without solution of the problem of compensation of
vacuum energy any model of dark energy cannot be considered as 
complete, though phenomenological suggestions~\cite{quint} 
can be quite useful.

In our paper~\cite{ad-mk-03} we considered a modification of the
model of Ref.~\cite{mr03} and as a result we were able to obtain
realistic cosmological solutions with dynamical compensation of
vacuum energy and non-compensated remnants or, better to say, 
an excessive energy density of $\phi$ being cosmological dark energy.
To our mind the mechanism described in \cite{ad-mk-03} can be 
considered as moderately successful in the sense that such solutions
indeed exist. However, their stability has not been studied and its
suspected absence may present a serious challenge to a possibility of 
simultaneous solution of both vacuum and dark energy problems in the 
considered frameworks. In this paper we study the stability of
the solutions presented in Ref.~\cite{ad-mk-03}
and find that they are unstable with respect to 
small perturbations. However, with some modification of the underlying
Lagrangian governing interaction of $\phi$ with gravity stability can
be achieved but at the expense of a realistic transition from radiation
domination stage to matter one.

The action of the considered model has the form:
\be
A= \int d^4 x\sqrt{g} \left[ - \frac{(R+2\Lambda)}{2} + 
\frac{ D_\mu\, \phi D^\mu \phi}{2 R^2} - U(\phi)
\right]
\label{A}
\ee
where the metric has the signature $(+,-,-,-)$, $D_\mu$ is the 
covariant derivative in this metric, and $g=-\det[\gmn]$. We took
the units such that $\mpl/8\pi =1$. Correspondingly
the cosmological constant $\Lambda$ is expressed through
the vacuum energy as $\rho_{vac} = \Lambda \mpl /8\pi= \Lambda$.
The explicit form of the potential $U(\phi)$ is not essential
because solutions tend to $\phi = \phi_0=const$ and only the 
magnitude of the derivative of the potential at this point,
$U'(\phi_0)$, determines asymptotic behavior of the solution.
Hereafter, we take $\phi_0 = 0$ for simplicity. 

In the cosmological Friedmann-Robertson-Walker (FRW) background 
equation of motion for spatially homogeneous field $\phi =\phi(t)$
takes the form:
\be 
\left( \frac{d}{dt} + 3 H \right)\left(\frac {\dot \phi}{R^2}
\right) + U' (\phi) =0.
\label{ddot-phi}
\ee
while the Einstein equations acquire additional terms related to 
$\phi$:
\be &&
 \rmn - \frac{1}{2} \gmn R 
 -\frac {D_\mu\phi D_\nu \phi}{R^2}
+ \frac{\left( D_\alpha \phi \right)^2}{2 R^2} 
\left({\gmn } + \frac{4\rmn}{R}\right)
\nonumber \\
&& -\gmn \left[ U(\phi) +\rho_{vac} \right]
+ 2\left(\gmn D^2 -D_\mu D_\nu\right) 
 \frac{\left( D_\alpha \phi\right)^2}{R^3} = 
 T_{\mu\nu},
\label{ein-eq}
\ee
where $T_{\mu\nu}$ is the energy-momentum tensor of the usual matter and 
$(D \phi)^2 \equiv D_\alpha \phi D^\alpha \phi$.

Taking trace over $(\mu-\nu)$ we obtain equation governing evolution
of the curvature scalar $R$:
\be
R - 3\, \frac {\left( D \phi \right)^2}{R^2} + 
4\left[ U(\phi) +\rho_{vac}\right] - 
6 D^2 \frac{\left( D_\alpha \phi\right)^2}{R^3} = T
\label{trace}
\ee
where $T = T^\mu_\mu$.
For the spatially homogeneous case the covariant D'Alambertian is
$D^2 = d^2/dt^2 + 3 H d/dt$ and the Hubble parameter, $H$, is 
expressed through $R$ as
\be
R = -6 \left( 2 H^2 +\dot H \right)
\label{r-of-h}
\ee

\begin{figure}[!t]
\begin{center}
\includegraphics[width=11cm]{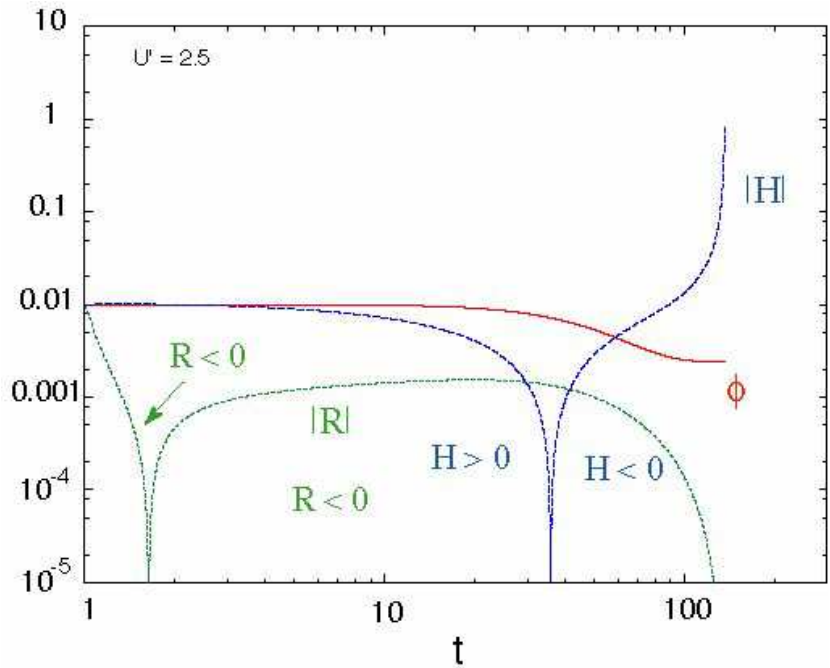}
\caption{Evolution of the scalar $\phi$, Hubble $H$ and the 
curvature $R$ for the potential $U'= 2.5$ and $\rho_m =0$. 
For initial conditions we take $\phi(0)=0.01$, $H(0) = 0.01$, 
$R(0) = -0.01$, $\phi'(0)= 10^{-5}$ and $R'(0) = 0$.}
\label{fig:collapse}
\end{center}
\end{figure}

\begin{figure}[!t]
\begin{center}
\includegraphics[width=11cm]{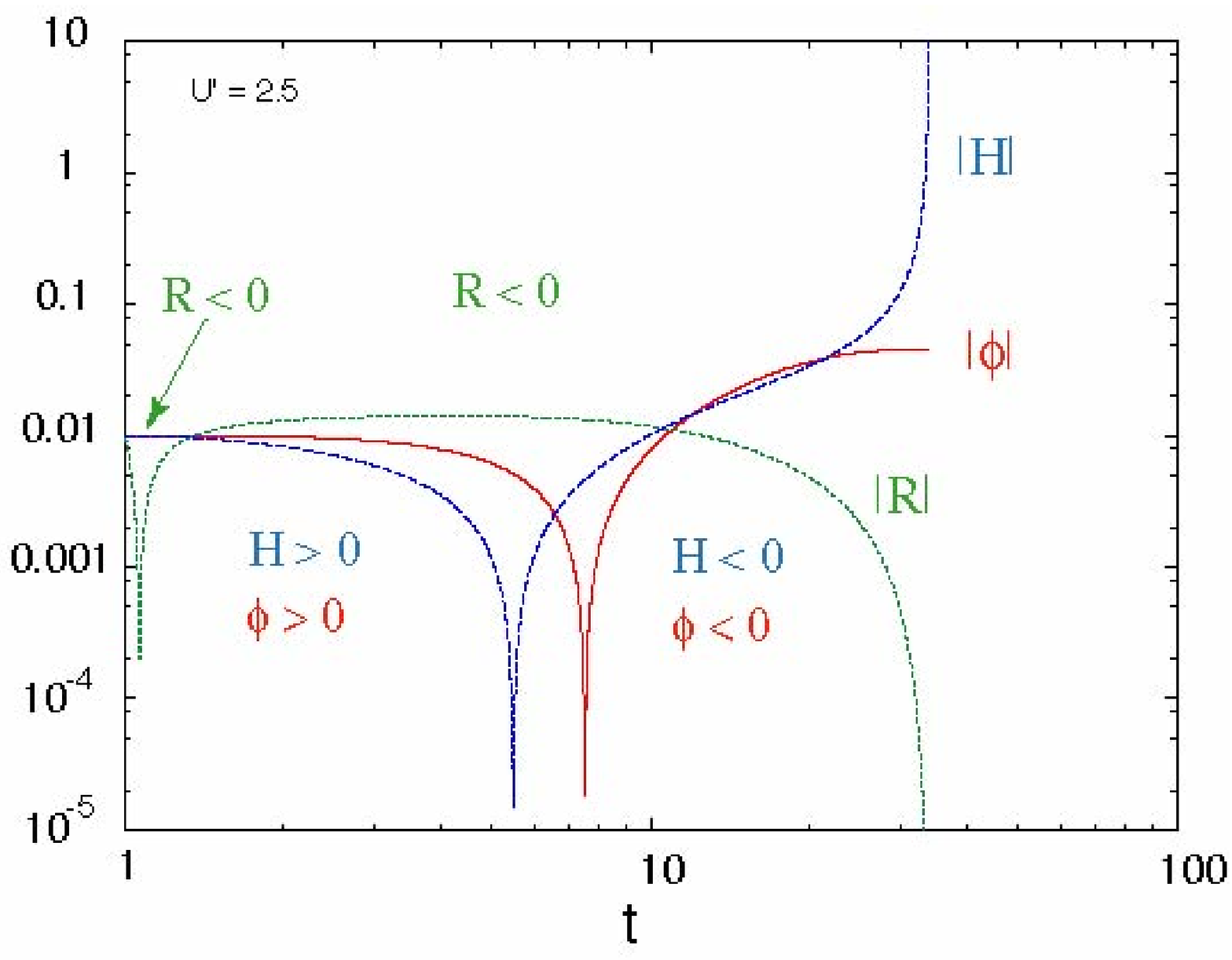}
\caption{Same as Fig.~\ref{fig:collapse} except for
$\rho_{m}(0)=1$.}
\label{fig:collap_w_m}
\end{center}
\end{figure}

\begin{figure}[!t]
\begin{center}
\includegraphics[width=11cm]{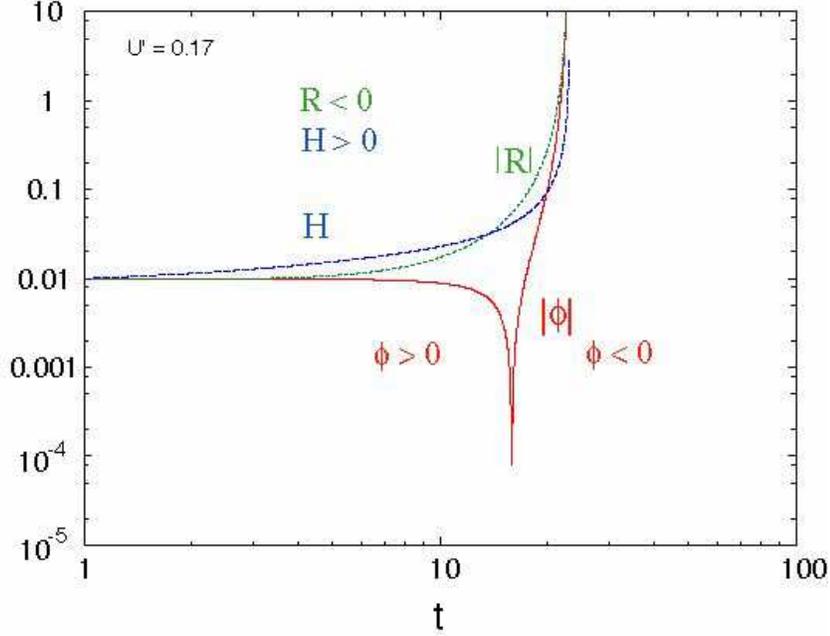}
\caption{Same as Fig.~\ref{fig:collapse} except for
$U'=0.17$.}
\label{fig:singular}
\end{center}
\end{figure}

As is found in Ref.~\cite{ad-mk-03}, these equations allow power law
solutions, $H=h/t$, $R=r/t^2$, and $[U(\phi) - \vacr] \sim 1/t^2$. An
interesting feature of this solution is that the term containing $D^2$ 
in Eq.~(\ref{trace}) is subdominant and can be neglected. However, for
the study of stability one, as is well known, must retain higher 
derivative terms in differential equation of motion. We checked 
numerically that the solution of equations 
(\ref{ddot-phi},\ref{trace},\ref{r-of-h}) quickly becomes singular with
$H$ and/or $R$ going to infinity in finite time. General solution
leads to a change of sign of $R$ from initially negative value to 
a positive one and to subsequent collapse ($H \rightarrow -\infty$)   
as shown in  Fig.~\ref{fig:collapse} for $\rho_m=0$ and
Fig.~\ref{fig:collap_w_m} for $\rho_m \neq 0$. 
On the other hand, when the potential $U$ is flatter, i.e. smaller $U'$,
the scalar field first changes sign and then all variable 
($\phi$,$H$ and $R$) blow up for $\rho_m =0$ 
(Fig.~\ref{fig:singular}).\footnote{
If there exists significant $\rho_m$ the solution behaves as seen 
in Fig.~\ref{fig:collap_w_m} even for small $U'$.}
In this way a catastrophic cosmological singularity is quickly approached.

To cure this unpleasant behavior we changed the Lagrangian (\ref{A})
in such a way that the solution of equations of motion for $\phi$ 
(\ref{ddot-phi}), $R$ (\ref{trace}) and $H$ (\ref{r-of-h}) 
does not lead to the catastrophic  collapse of the universe 
for any sign of the curvature $R$. This may be done if the Lagrangian 
is allowed to depend upon the absolute value of $R$.\footnote{
This modification cannot cure the blow up of $\phi$,$H$ and $R$ in
the case of small $U'$ and $\rho_m$ since this instability arises 
without change of the curvature sign.}
If we modify the kinetic term of the $\phi$-field as
\be
\frac{(D \phi)^2}{R^2} \rar - \frac{(D \phi)^2}{R\,|R|}
\label{absr} 
\ee
then $|H|$ would not infinitely rise for any sign of $R$.
Such non-analytical terms in the Lagrangian are rather unusual but may
be allowed for a toy model. One may have slightly better form for
non-analytic terms as e.g. $\sqrt{R^2+(D_\alpha \phi)^2}$.
To avoid negative sign of the expression under square root the solution
should ensure $R^2>|(D\phi)^2|$, or if it is not so the expression could
be properly modified e.g. by changing $(D\phi)^2$ into $(D\phi)^4$.

It is convenient to introduce additional notations:
\be
 z=\frac{\dot\phi}{R|R|}\,\,\,{\rm and}\,\,\, 
w=\frac{(\dot\phi)^2}{|R|^3}
\label{not}
\ee
In terms of these new functions and in spatially homogeneous case we 
obtain the following system of equations governing cosmological
evolution in the model under consideration:
\be
&&\dot\phi- \frac{|w|\,w}{z^3} = 0,
\label{dot-phi}
\\
&&\dot z + 3H z -U'(\phi) = 0,
\label{dot-z}
\\
&&\dot H + 2H^2 +R/6 = 0,
\label{dot-h}
\\
&&6 D^2 w + 3\frac{w^2}{z^2} + \frac{|w|}{z^2} + 4\frac{|w|}{w}\left[
U(\phi) +\vacr\right] +\frac{|w|}{w}\,T =0.
\label{eqns}
\ee
The evolution of the trace of the energy-momentum tensor of the 
usual matter is determined by its equation of state. For relativistic 
matter $T=0$, while for non-relativistic matter:
\be
\dot T = 3HT
\label{dotT}
\ee

\begin{figure}[!t]
\begin{center}
\includegraphics[width=11cm]{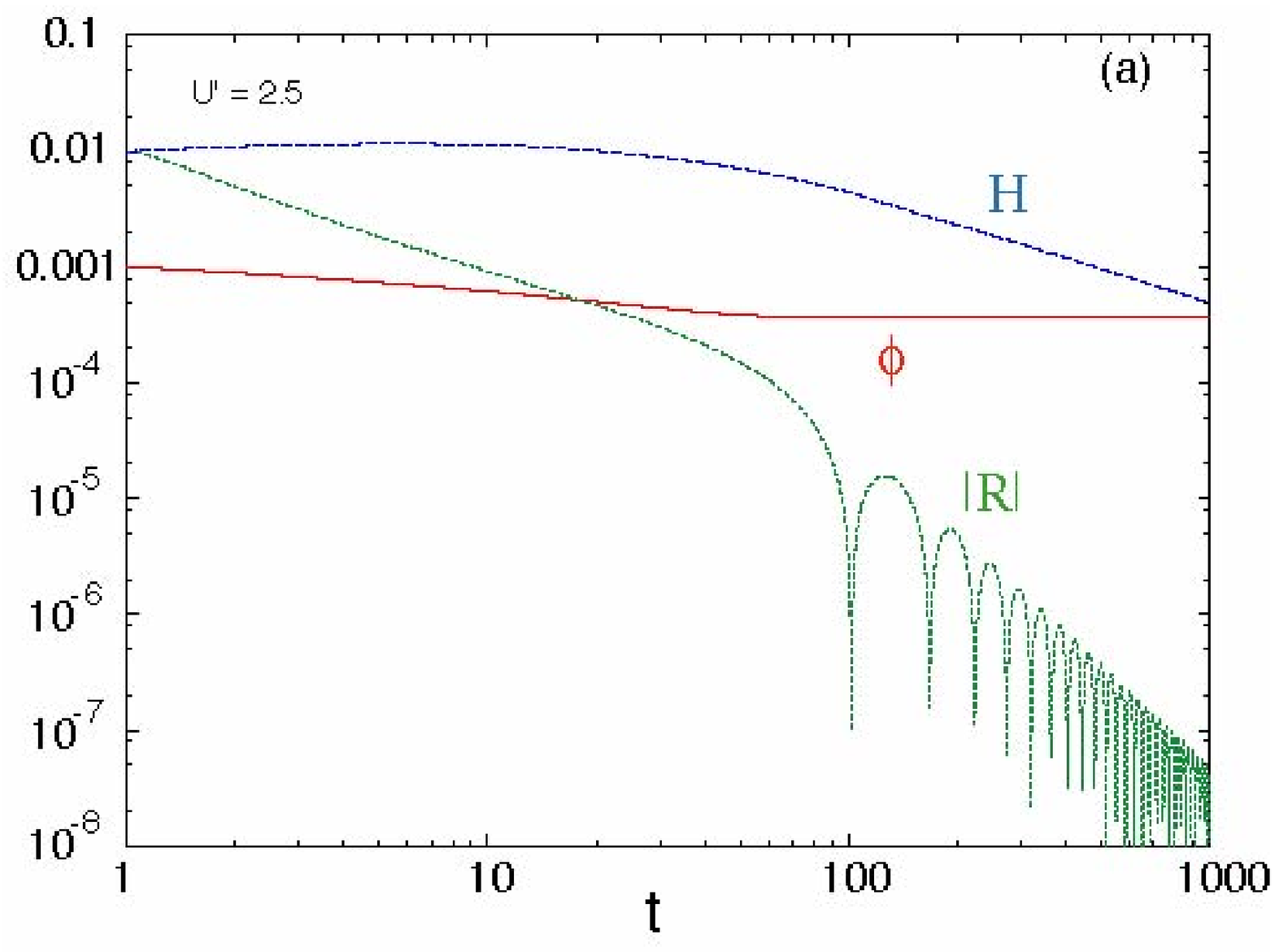}
\includegraphics[width=11cm]{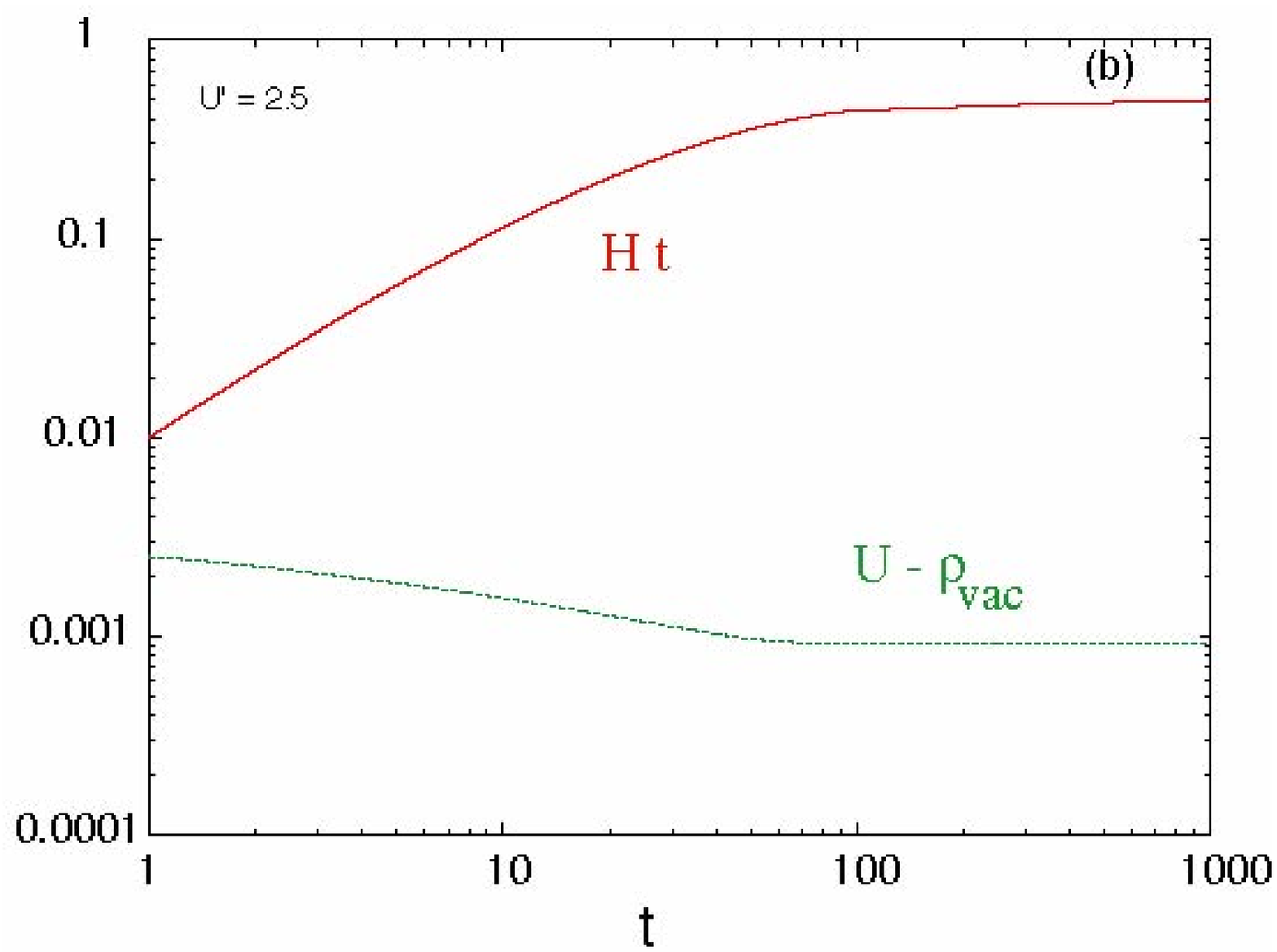}
\caption{(a) Evolution of the scalar $\phi$, Hubble $h$ and the 
curvature$R$ for the modified kinetic term Eq.(\ref{absr}).
We take  $U'= 2.5$ and $\rho_m =0$. The initial values are
$\phi(0)=0.001$, $H(0) = 0.01$, $R(0) = -0.01$, $\phi'(0)= 10^{-4}$ 
and $R'(0) = 0$. (b) Evolution of $h=Ht$ and $U-\rho_{vac}$.}
\label{fig:osc_wo_m}
\end{center}
\end{figure}

\begin{figure}[!t]
\begin{center}
\includegraphics[width=11cm]{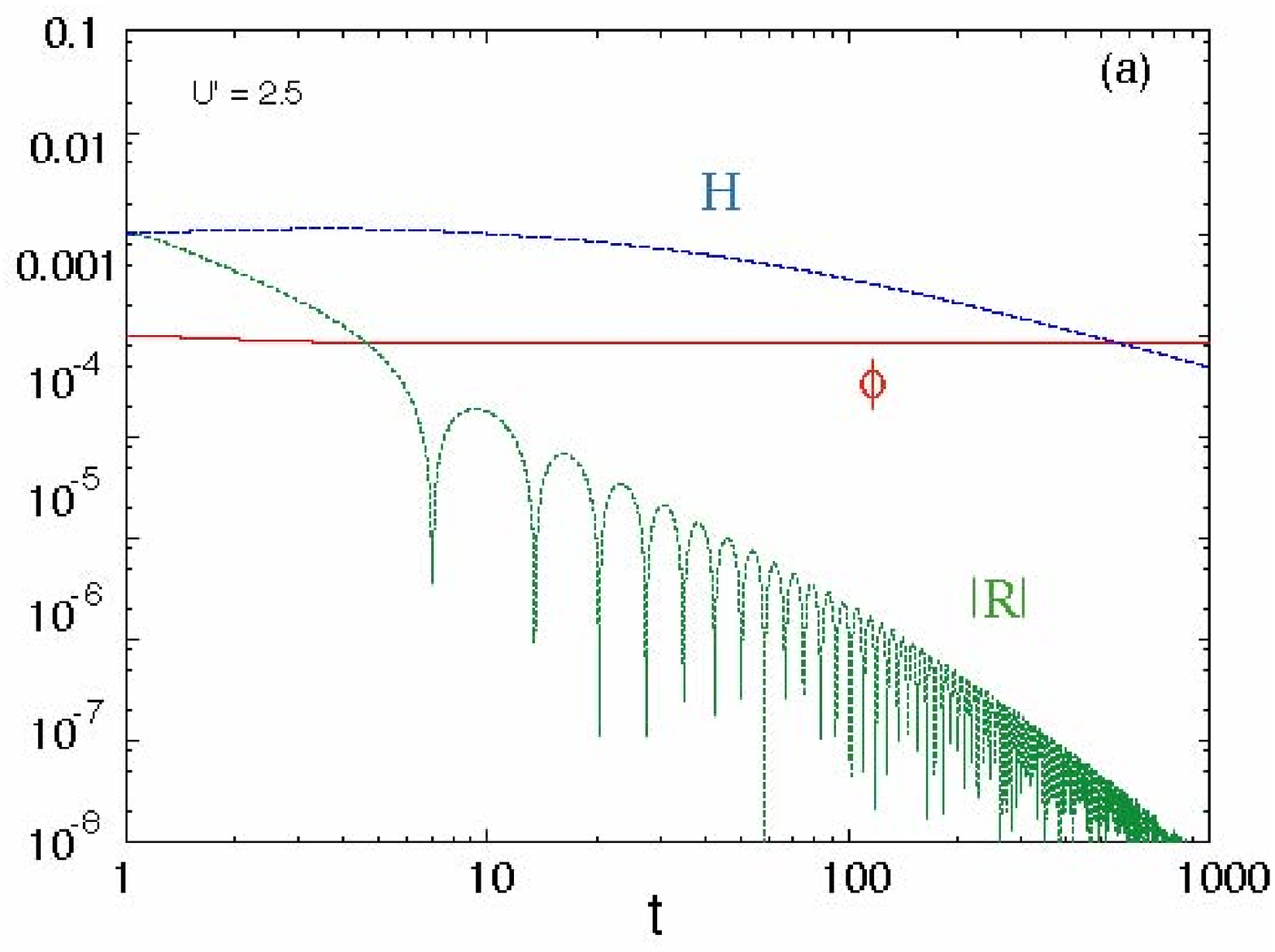}
\includegraphics[width=11cm]{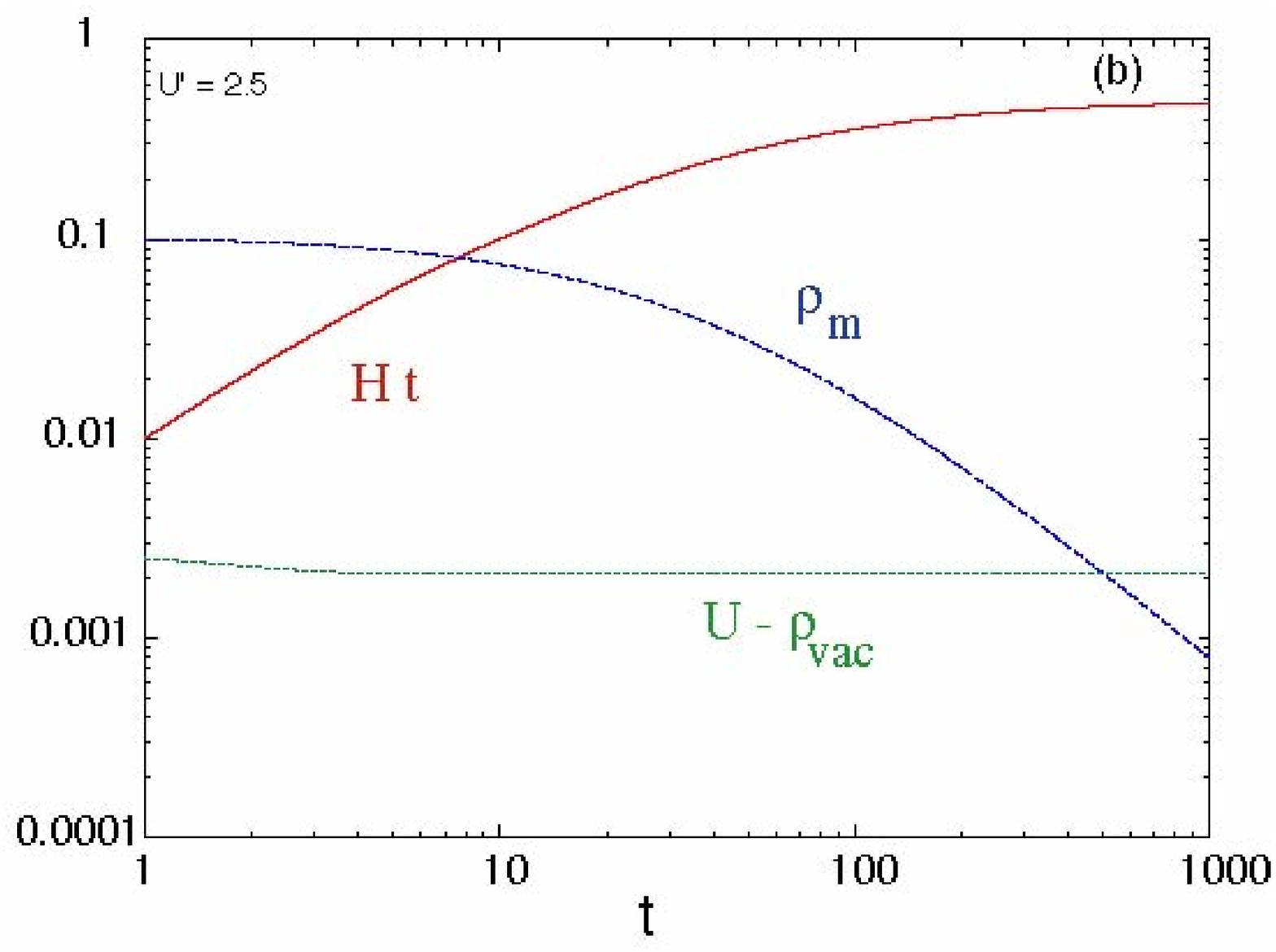}
\caption{Same as Fig.~\ref{fig:osc_wo_m} except for
$\rho_{m}(0)=0.1$.}
\label{fig:osc_w_m}
\end{center}
\end{figure}

The system of equations (\ref{dot-phi}-\ref{dotT})
has been solved numerically, see 
figs.~\ref{fig:osc_wo_m} and \ref{fig:osc_w_m}.
We have found, for more or less general initial conditions, 
that the curvature, $R$, oscillates around zero with an 
increasing frequency and with the amplitude decreasing somewhat 
faster than $1/t^2$. Possibly particle production by quickly 
oscillating curvature would make the magnitude of $R$ to decay even  
faster. The Hubble parameter tends to $H=1/(2t)$, as 
is the case of cosmology dominated by relativistic matter. The
compensating scalar field $\phi$ tends to a constant value oscillating
around the latter. In the case that non-relativistic matter is absent 
and thus, $T=0$ the model corresponds to reasonable Friedmann cosmology
with the usual expansion law, $a(t)\sim t^{1/2}$. However, if 
non-relativistic matter is present and $T$ is non-zero it 
would start to
dominate the total matter energy density, $T=\rho_m\sim a^{-3}$, 
while the expansion regime remains relativistic, $a\sim t^{1/2}$.

Still the solutions found in Ref.~\cite{ad-mk-03} also exist but to 
``hit'' them a fine-tuning of initial conditions is necessary. Moreover,
these solutions are unstable and small fluctuations around them have
one (out of 5) rising mode. It is interesting that if one neglects
$D^2$-term in Eq.~(\ref{trace}), as we have done in Ref.~\cite{ad-mk-03},
the ``good'' solutions found there
are stable but the transition from
relativistic to non-relativistic regime encounters instability.

So only a moderate success can be reported. The
model presented above allows for the transformation of de Sitter
cosmological solution, driven by vacuum energy,
into the Friedmann one but the expansion regime
is always relativistic with $H=1/(2t)$ independently of the matter 
content. The matter dominated (MD) solution with $H=2/(3t)$
is also possible but such
solution is unstable, small fluctuations around it would drive the 
system back to radiation dominated (RD) regime. Moreover, starting 
from RD regime, assuming that $T$ is small, we have not found
any solution which allows for the change to MD regime when $T$ begins
to dominate matter density. Despite these shortcomings, one may hope
that the resolution of these problems may be feasible with some
modification of the Lagrangian describing interaction of the compensating
field with gravity. Maybe such constructions resembles epicycles in
Ptolemaic astronomy but there surely was some truth in the latter. 
Anyhow, adjustment mechanism seems to be the 
only one that may possibly solve 
simultaneously both problems of 100 orders of magnitude compensation
of vacuum energy and to explain the non-compensated remnant in 
cosmological energy density (dark energy) which contributes a relative
fraction of order of unity into cosmological energy density at any time
of the universe history.

\bigskip

\noindent
{\bf Acknowledgment} 
We thank S. Mukoyama for critical comments on our previous 
paper~\cite{ad-mk-03} related to stability of the solutions.
A.D. Dolgov is grateful to the Research Center for
the Early Universe of the University of Tokyo for the hospitality during
the time when this work was done.

\end{document}